# Doubling of the superconducting transition temperature in ultra-clean wafer-scale aluminum nanofilms

Ching-Chen Yeh[1,*], Thi-Hien Do[2,*], Pin-Chi Liao[1], Chia-Hung Hsu[3], Yi-Hsin Tu[4], Hsin Lin[5], T.-R. Chang[4], Siang-Chi Wang[1], Yu-Yao Gao[2], Yu-Hsun Wu[2], Chu-Chun Wu[2], Ivar Martin[6], Sheng-Di Lin[2,✉], Christos Panagopoulos[7,✉], Chi-Te Liang[1,8,✉]

[1]*Department of Physics, National Taiwan University, Taipei 106, Taiwan*
[2]*Institute of Electronics, National Yang Ming Chiao Tung University, Hsinchu 300, Taiwan*
[3]*National Synchrotron Radiation Research Center, Hsinchu 300, Taiwan*
[4]*Department of Physics, National Cheng Kung University, Tainan 701, Taiwan*
[5]*Institute of Physics, Academia Sinica, Taipei 115, Taiwan*
[6]*Materials Science Division, Argonne National Laboratory, Argonne, Illinois 08540, USA*
[7]*Division of Physics and Applied Physics, School of Physical and Mathematical Sciences, Nanyang Technological University, 21 Nanyang Link 637371, Singapore*
[8]*Center for Quantum Science and Engineering, National Taiwan University, Taipei 106, Taiwan*

*These authors contributed equally to this work.
✉e-mails: sdlin@mail.nctu.edu.tw, christos@ntu.edu.sg, and ctliang@phys.ntu.edu.tw

Superconducting properties of thin films can be vastly different from those of bulk materials. Seminal work has shown the critical temperature $T_c$ of elemental superconductors decreases with decreasing film thickness when the normal-state sheet resistance is lower than the quantum resistance $h/(4e^2)$. Sporadic examples on disordered films, however, hinted an enhancement in $T_c$ although, structural and strain characterization was not possible since samples were prepared on a cold substrate *in situ*. To clarify the role of reduced dimensionality and disorder on the superconducting properties of thin films we employed molecular beam epitaxy to grow wafer-scale high-quality aluminum (Al) nanofilms with normal-state sheet resistance at least 20 times lower than $h/(4e^2)$ and investigated their electronic and structural properties *ex situ*. Defying general expectations, $T_c$ *increases* with decreasing Al film thickness, reaching 2.4 K for 3.5-nm-thick Al film grown on GaAs: twice that of bulk Al (1.2 K). DFT calculations indicate surface phonon softening impacts superconductivity in pure ultra-thin films, offering a new route for materials engineering in two dimensions.

Superconductivity in thin films of conventional superconductors is anticipated to be suppressed with decreasing sample thickness, leading to a superconductor-insulator transition (*1,2*). Early studies on aluminum (Al) however, reported unexpected enhancement of $T_c$ in thin films, which were polycrystalline, granular, or included Ge or $Al_2O_3$ (*3-7*). The apparent contrast on the effect of thickness on superconductivity (*1-20*) challenged our understanding of disorder-induced localization of Cooper pairs, Coulomb screening, and the generation and unbinding of vortex−antivortex pairs in low dimensions. The intrinsic trend of $T_c$ with film thickness has remained an enigma for several decades with the discussion rekindled recently, following results on single-layer FeSe showing an unprecedentedly high $T_c$ compared to the bulk counterpart (*10-13*).

Theoretically, an increase of $T_c$ in thin films was predicted by Blatt and Thompson (*21*). The basic effect relies on an enhanced electronic density of states due to electronic confinement. Here, each transverse electronic sub-band has approximately a constant density of states, with the total exceeding the bulk density of states at the Fermi level, thus enhancing $T_c$. The effect becomes particularly noticeable when only a small number $n$ of transverse sub-bands is occupied; there the relative enhancement of DOS scales as $1/n$. Subsequent detailed calculations (*22*) took into account the renormalization of the chemical potential in thin films and pointed out that the strength of the confining potential matters ($T_c$ is enhanced for strong confinement and may be suppressed for weak confinement, which points to the importance of the substrate). At a similar level, it was also shown that the transverse confinement of phonons leads to additional features in $T_c$ (*23*). These theoretical models provide a clear motivation for studying superconductivity in thin films and guide more detailed microscopic calculations.

To clarify the effect of film-thickness on $T_c$, we have grown a series of pure Al ultra-thin films on three different substrates using molecular beam epitaxy (MBE) and investigated both the electronic and structural properties. In addition to being a well-studied conventional BCS superconductor, Al is the most abundant metallic element in the Earth’s crust and employed in a plethora of applications. These include heat sinks (*24*), interconnects (*25*), plasmonics (*26*), superconducting detectors (*27*), metallic supercurrent field-effect transistors (*28*), and quantum computation, science and technology (*29*). Identifying the mechanism governing $T_c$ in wafer-scale pure nanofilms would allow to explore superconductivity in the ultra-thin film limit and to develop a recipe for quality control in the many applications of Al. Moreover, enhancing $T_c$ in relevant nanostructures may lead to the much desired, lower quasi-particle concentration at mK temperatures, substantially increasing the coherence time of a superconducting qubit (*30*).

Figures 1 A and B depict the sheet resistance $R$ for Al films of different thicknesses grown on GaAs and on sapphire, respectively. $T_c$ is the temperature at which the resistance ratio crosses 50% of $R/R_N$. Plots of the ratio of sheet resistance to the normal-state sheet resistance as a function of temperature $R/R_N(T)$ are shown in figures S1A and B. $T_c$ decreases with increasing film thickness $d$ (Fig. 1C). For Al films in the range 3.5 nm $\leq d \leq$ 10 nm, $T_c$ is higher for films grown on GaAs and converges to the value for bulk Al (1.2 K) for thicker films ($d =$ 20 nm) (Fig. 1C).

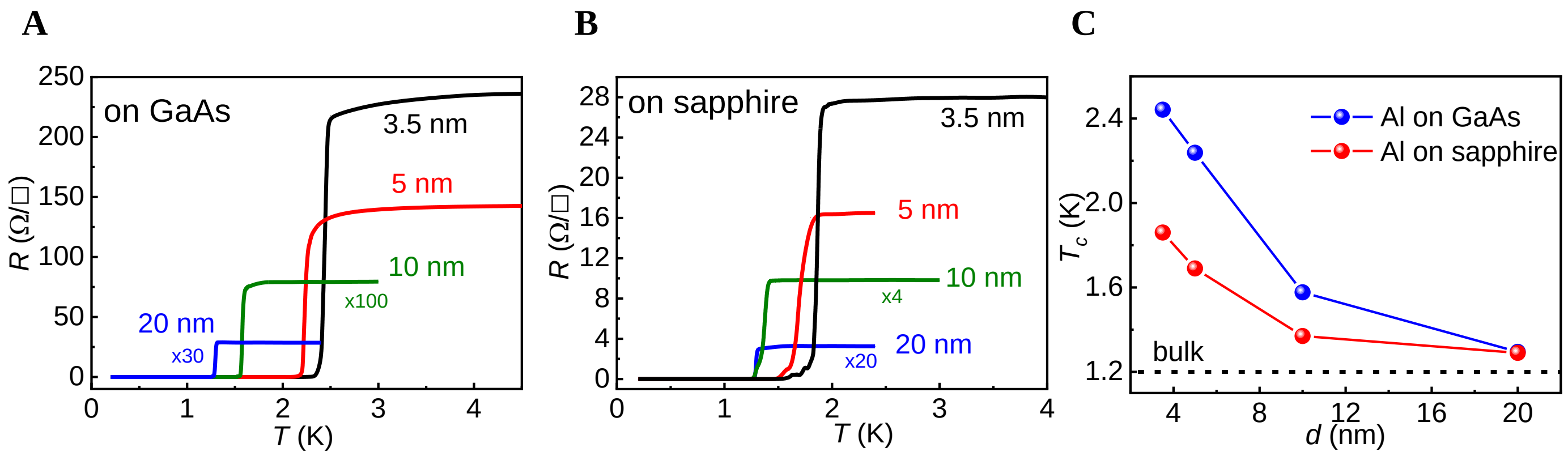

**Fig. 1. Resistive transitions of Al nanofilms.** Sheet resistance $R$ as a function of temperature for (**A**), Al nanofilms of different thicknesses $d$ grown on GaAs. (**B**) Al nanofilms of different thicknesses grown on sapphire. **(C)** Critical temperature $T_{\mathrm{c}}$ for Al films of different thicknesses grown on GaAs and on sapphire.

To further probe the observed enhanced superconductivity in the 3.5-nm-thick Al nanofilms grown on GaAs and sapphire, we performed electrical measurements on these two sets of films, as well as a counterpart film grown on Si (Fig. S1C). Figure 2A depicts the current-voltage ($I$-$V$) characteristics of a 3.5-nm-thick Al film grown on GaAs. Measurements taken on counterparts grown on sapphire and Si are shown in Figs. S2A and S2B.

From $V \sim I^{\beta}$, when $\beta \approx 3$ (*31*, *32*) the corresponding Berezinskii-Kosterlitz-Thouless (BKT) transition (*33*, *34*) temperatures $T_{\mathrm{BKT}}$ for the 3.5-nm-thick Al films on GaAs (Fig. 2B), sapphire, and Si (insets of Figs. S2A and S2B) are 2.401 K, 1.824 K, and 2.057 K, respectively. These values agree with numbers determined from resistivity measurements (Fig. 1C and Fig. S1C). As expected, the difference between $T_{\mathrm{BKT}}$ $(V \sim I^{3})$ and $T_{\Phi}$ $(V \sim I^{1})$ decreases in thicker films and $T_{\Phi} > T_{\mathrm{BKT}}$ in the ultra-thin limit (Fig. 2C), consistent with studies on Pb (*32*), albeit the increasing $T_{\mathrm{c}}$ with decreasing film thickness in our ultra-thin Al films. Here, $T_{\Phi}$ is the intrinsic transition temperature due to Cooper pair breaking. The high quality of our Al nano-films is further reflected in the small differences between the transition temperatures determined from $R = 0.9R_{\mathrm{N}}$ and $R = 0.1R_{\mathrm{N}}$ (Figs. S3A, S3B, and S3C).

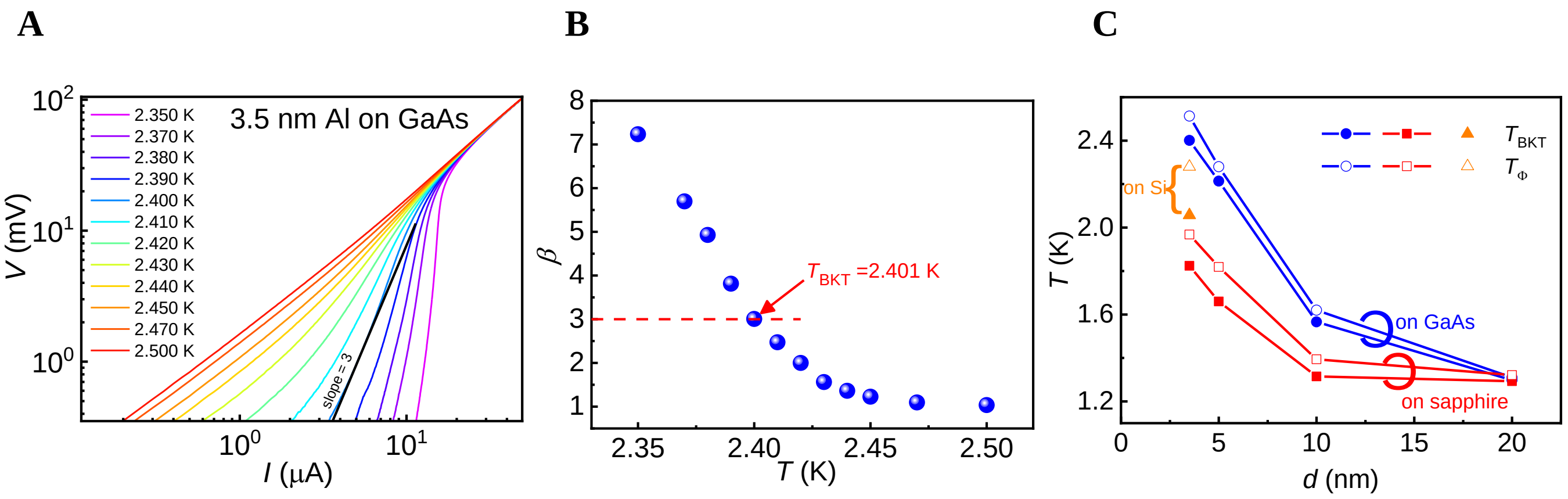

**Fig. 2. Current-voltage measurements and comparison of transition temperatures.** (**A**) $I$-$V$ characteristics obtained at various temperatures for a 3.5-nm-thick Al film grown on GaAs. Panel **(B)** shows the exponent $\beta$ ($V{\sim}I^{\beta}$) for the same film. **(C)** $T_{\mathrm{BKT}}$ and $T_{\Phi}$ ($V{\sim}I^{1}$) for Al on GaAs and on sapphire at various thicknesses. The plot includes data for a 3.5-nm-thick Al film grown on Si.

Using out-of-plane magneto-sheet resistance data $R^{\perp}(H)$ obtained on 3.5-nm-thick Al films at different temperatures (Figs. S4A, S4B, and S4C), we identify $H_{\mathrm{c}}$ as the magnetic field where $R^{\perp}$ crosses 50% of its residual normal-state value. From $H_{\mathrm{c}}(T) = H_{\mathrm{c}}(T=0)[1-(T/T_{\mathrm{c}})^2]$, we calculate the Landau-Ginzburg coherence length $\xi(T=0)$ (28 nm, 78 nm, and 44 nm for Al on GaAs, sapphire, and Si, respectively). Combined with the observed BKT transitions ($V{\sim}I^{3}$) (Figs. S5A – 5F), these numbers (considerably longer than the film thickness) add credence to two-dimensional superconductivity.

Next, we discuss the structural properties of our samples. Atomic force microscopy (AFM) on 3.5-nm-thick Al films reveals root-mean-square (RMS) roughness of 0.265 nm, 0.268 nm, and 0.316 nm for Al nanofilms grown on GaAs, sapphire, and Si, respectively (Figs. S6 A-C). X-ray diffraction (XRD) $2\theta$-$\theta$ scans along the surface normal, in units of substrate reciprocal lattice unit (rlu) depict weak and broad albeit pronounced Al (111) peaks for films grown on Si and sapphire substrates (Figs. 3B and 3C). Whereas for films grown on GaAs, we observed a weaker and broader Al (111) peak on a strong background (Fig. 3A). The broad peak widths are ascribed mainly to the small film thickness.

We also performed lateral radial scans for Al films grown on GaAs, sapphire, and Si substrates (Figs. S7 A-C). For 3.5-nm-thick films on sapphire and Si, the Al $(2\bar{2}0)$ reflection exhibits a lattice constant close to bulk Al. There is also a peak (peak B) in addition to the Al $(2\bar{2}0)$ reflection (peak A) in the lateral radial scan along [110] in GaAs. Furthermore, the corresponding inter-planar spacing (1.43 Å) of peak A agrees with that of the bulk Al $(2\bar{2}0)$ reflection. Analysis shows peak B has a different structure from peak A and is associated with the region nearer the surface (Fig. S8). All XRD results indicate our Al (111) films were grown epitaxially on GaAs (001), c-plane (0001) sapphire, and Si (111) substrates.

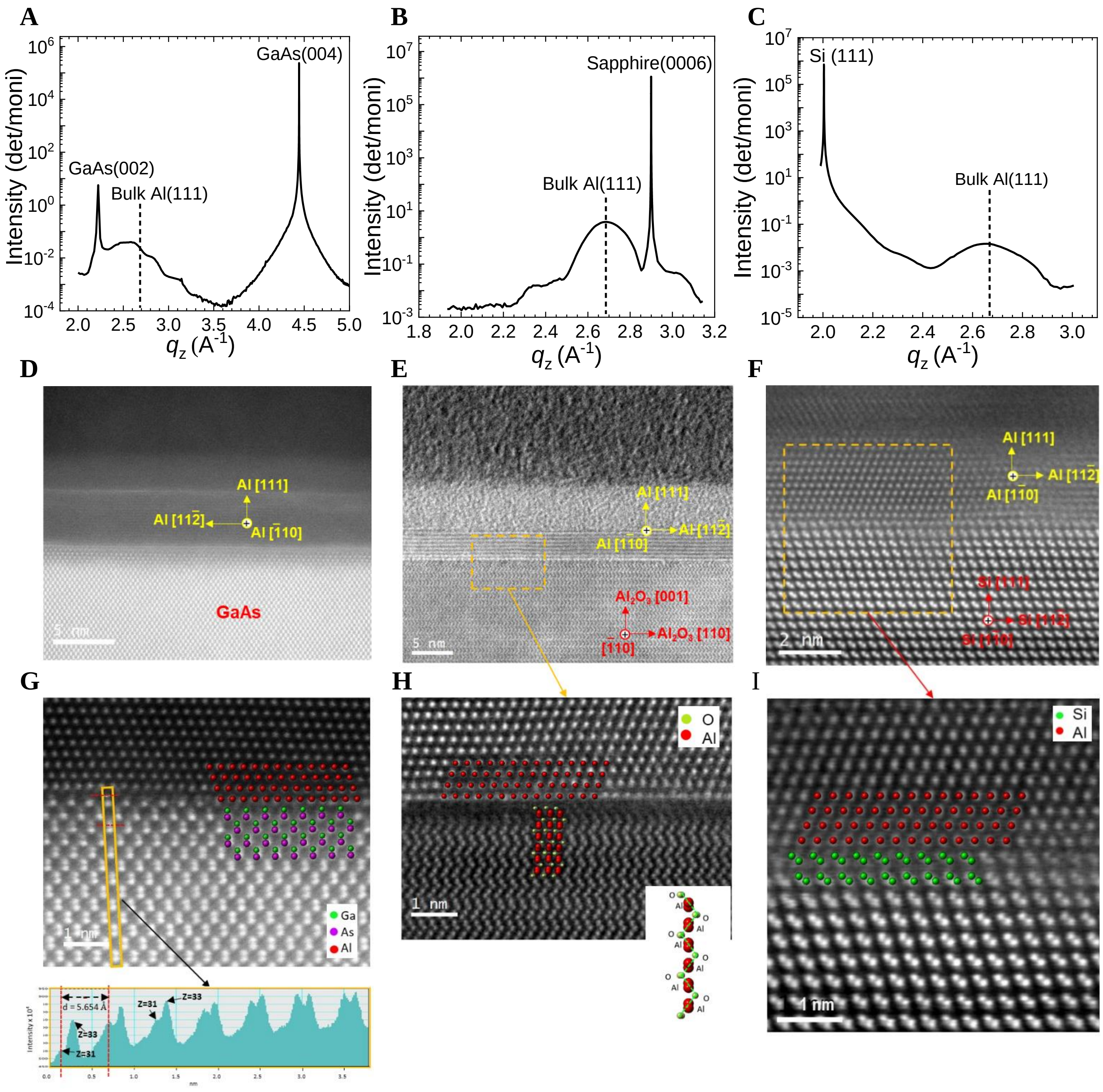

**Fig. 3**. **Structural and material characterization of 3.5-nm-thick Al films.** XRD radial scans along surface normal of 3.5-nm-thick Al films grown on **(A)** GaAs, **(B)**, sapphire, and **(C)** Si, respectively. The corresponding $q_z$'s for bulk Al (111) are indicated by the dotted lines. **(D)-(F)** Top panel: Cross-sectional HAADF images of 3.5-nm-thick Al films grown **(D)** on GaAs, **(E)** on sapphire, and **(F)** on Si, respectively. **(G)-(I)**: enlarged selected areas of 3.5-nm-thick Al films.

Microstructure studies using high-resolution TEM/STEM add credence to the epitaxial growth of our Al films. Figures 3 D-F depict cross-sectional high-angle annular dark field (HAADF) images revealing high-quality ultra-thin epitaxial Al films on GaAs, sapphire, and Si. Corresponding fast-Fourier-transform patterns (not shown) indicate an orientation consistent with the above mentioned XRD results. Furthermore, the interfaces are sharp with no interlayer mixing between Al films and substrates.

Using enlarged atomic-resolution HAADF images of the selected area in Figs. 3H, 3I, we resolved the atomic structure of Al films and substrates, revealing interfacial bonding of Al-O and Al-Si for the Al films grown on sapphire and Si substrates, respectively. The enlarged image of Fig. 3G depicts the Al/GaAs interface and dumb-bell structure of Ga/As atomic columns. We note, $Z_{Ga}$= 31 and $Z_{As}$ = 33. We verified the Ga/As atomic columns, using the STEM intensity profile across the dumb bell. As shown in the inset of Fig. 3G, both Ga and As atomic columns can be resolved. The terminating plane is Ga, in agreement with the surface treatment condition before growth (Ga-rich surface). Hence, the interfacial bonding is Al-Ga.

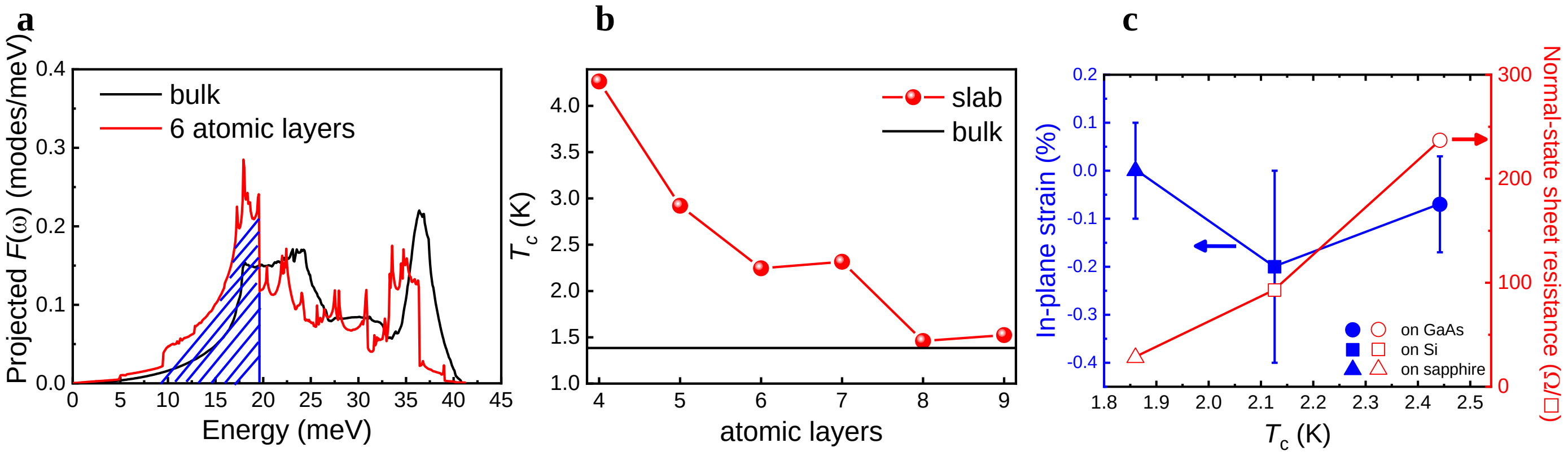

**Fig. 4. DFT calculations and the experimentally determined dependence of $T_c$ on the normal-state sheet resistance and in-plane strain. (A)** Projected phonon density of states for the top-most surface atom. The black and red curves correspond to bulk and 6-layer Al thin film, respectively. The hatched regions indicate phonon-softening in thin Al films in the low energy regime. **(B)** Thickness dependence of $T_c$. **(C)** $T_c$ versus in-plane strain and low-$T$ normal-state sheet resistance for the 3.5-nm-thick Al films on the three different substrates explored in this study. The highest normal-state sheet resistance in our films is 20 times lower than the quantum resistance $h/(4e^2)$.

Having characterized the structural and electronic properties of our Al films, we turn our attention to the physical mechanism leading to the enhancement of $T_c$ with decreasing film thickness. We determined the electronic structure, phonon density of states (DOS), and electron-phonon coupling properties via first-principles calculations (Figs. S9 A-I). Fig. S9F depicts the isotropic Eliashberg spectral function $\alpha^2F(\omega)$ for bulk Al and a 6-layer Al thin film. Compared to bulk Al, the highest frequency peak (36.5 meV) is suppressed and the intensity in the lower-energy state is increased in the 6-layer Al thin film. As a result, the electron-phonon coupling strength $\lambda = 2\int \alpha^2F(\omega)\omega^{-1}d\omega$ increased by about 13%. The phonon DOS $F(\omega)$ exhibits a similar form to $\alpha^2F(\omega)$, as shown in Fig. S9G. Since $\alpha^2F(\omega)$ is related to $F(\omega)$, we can infer the trend of $\alpha^2F(\omega)$ by studying $F(\omega)$ (Fig. S9H). Based on layer-resolved decomposition, we attribute the low-energy phonon DOS mainly to surface phonons (hatched regions in Fig. 4A and Figs. 9F and 9G). Our DFT calculations show that the spacing of Al atoms on the film surface is slightly higher than for bulk Al. This modulation will soften the surface phonon mode and increase the intensity of the low-energy phonon DOS.

The calculated electronic density of states $N(E_F)$ as a function of Al layer thickness is depicted in Fig. S9H. Both $\lambda$ and $T_\mathrm{c}$ *increase* with decreasing film thickness (Fig. S9 I and Fig. 4B), consistent with our experimental results (Fig. 1C). Furthermore, Fig. 4C shows that $T_\mathrm{c}$ *increases* with increasing normal-state sheet resistance, in contrast to earlier studies (*1*,*2*). Hence, the normal-state sheet resistance is not the dominant factor determining $T_\mathrm{c}$ for the thinnest films. Similarly, in-plane strain does not appear to govern the trend observed in $T_\mathrm{c}$ for the 3.5-nm-thick Al nanofilms grown on GaAs, sapphire, and Si (Fig. 4C). Neither do disorder (determined by the low-temperature normal-state sheet resistance and the residual-resistance ratio (*RRR*) $R_{300\,\mathrm{K}}/R_\mathrm{N}$ of the three 3.5-nm-thick Al films (Fig. S11)) and twin structures (seen as alternating peak strengths in a thick film grown on Si, and the pronounced six peaks in thick films on sapphire and on GaAs: Fig. S12), or strain (as shown in Table S1).

We conclude the increase in $T_\mathrm{c}$ is due to enhanced surface phonon softening and increased electronic density of states with decreasing film thickness. Our experimental results are in line with the work by Blatt and Thompson (*21*) where an enhanced electronic density of states due to confinement leads to an increase in $T_\mathrm{c}$. We also note that for the 3.5-nm-thick Al films, sapphire has the largest bandgap (10.0 eV) amongst the studied substrates, yet the enhanced $T_\mathrm{c}$ is the lowest compared to counterparts grown on Si and GaAs.

Next, we comment on the progressive increase in $T_\mathrm{c}$ for the 3.5-nm-thick Al films grown on sapphire, Si, and GaAs, respectively. Earlier works (Fig. S13) suggest the phonon DOS for GaAs is highest in the low-frequency limit but low and broad for sapphire (*35*-*37*). Since $\lambda \sim \alpha^2 F(\omega)/\omega$, assuming the electronic contribution from Al and the electron-phonon coupling from the substrates are similar, the low-frequency end of the phonon DOS may contribute to the rise in $\lambda$ and $T_\mathrm{c}$ with decreasing film thickness, resulting in the observed trend in $T_\mathrm{c}$ for the 3.5-nm-thick Al grown on different substrates.

Compared to earlier studies (*38*) the low temperature resistivities $\rho$ of our Al films are significantly lower (for the relevant film thickness) (Fig. S13A). Namely, for the 3.5-nm-thick and 5-nm-thick Al films grown on GaAs, our values for $\rho$ are closer to Ref. (*9*), and lower than in Refs. (*3*), (*9*), and (*38*) for the rest of our Al films. Also, the sheet resistances of our films are always lower than in earlier reports (Fig. S13 B). Moreover, the scaling $d \times T_\mathrm{c} = AR^{-B}$, does not hold for our films (Figure S13 C) (*39*). These separate findings correlate the primary impact of the substrate on the superconducting properties of Al epitaxial ultra-thin films, calling for further investigation of this aspect.

In summary, we report extensive transport and structural studies on high-quality wafer-scale epitaxial Al nanofilms grown by MBE. $T_c$ increases with decreasing film thickness, in contrast to earlier studies on Bi, Pb, and Al grown on cold substrates (20 K) (*1*, *38*), Pb (*32*), and Nb (*40*). DFT calculations suggest the observed increase in $T_c$ is due to surface phonon softening and enhanced electronic density of states with decreasing film thickness. Furthermore, $T_c$ for the thinnest Al nanofilms (3.5 nm) increases when grown on sapphire, Si, and GaAs; substrates commonly employed in plasmonics, Si-based quantum computation and information and Si CMOS technology, and high-frequency devices, respectively. Advances in the growth of pure ultra-thin films of elemental BCS superconductors in general, are therefore, expected to increase $T_c$ compared to bulk counterparts. This development impacts not only the long-standing investigation of superconductivity in the very thin limit but also the optimization of the normal and superconducting state properties in low-temperature technology.

## Methods

### Film growth by MBE

Ultra-thin epitaxial Al films were grown in a Varian Gen-II III-V solid-source MBE system (*41*, *42*). Si, GaAs, and sapphire substrates were pretreated prior to depositing the Al films. Before loading the substrate into the MBE chamber, the phosphorus-doped Si (111) wafer was treated as described previously (*42*). All Si, GaAs, and sapphire wafers were loaded into the MBE chamber and baked at 200 $^{o}$C at exit/entry chamber for 8 hours to remove moisture from the surface. Subsequently, Si and GaAs were heated to 400 $^{o}$C for 5 hours, while sapphire was heated to 500 $^{o}$C for 5 hours to remove any organic residues in the preparation chamber. In the growth chamber, Si wafer was heated to 630 $^{o}$C and cooled in the ultra-high vacuum chamber to less than 0 $^{o}$C before the Al nanofilm was deposited. Prior to depositing Al, a semi-insulating GaAs (001) substrate was heated to 600 $^{o}$C for 20 min to remove the native oxide, and a 200-nm-thick undoped GaAs buffer layer was then grown at 580 $^{o}$C. Following that, the GaAs substrate was heated for 3 minutes to 600 $^{o}$C without As flux to develop a Ga-rich surface. The GaAs wafer was subsequently cooled in the ultrahigh vacuum chamber to less than 0 $^{o}$C. The Al nanofilm was deposited after the residual As in the growth chamber was pumped out in a background pressure below $2 \times 10^{-10}$ Torr. The sapphire (0001) substrate was heated to 650 $^{o}$C for 1 hour to remove the native oxide, and then cooled to less than 0 $^{o}$C. The Al nanofilm was subsequently grown on the treated sapphire substrate.

### AFM, XRD and TEM/STEM measurements

Surface morphology of our samples was examined by AFM (Bruker Edge) with the tapping mode. The crystal structure was characterized using synchrotron based XRD techniques at beamline TPS 09A of National Synchrotron Radiation Research Center (NSRRC) in Taiwan. Cross-sectional transmission electron microscopy (TEM) and scanning TEM (STEM) specimens were prepared by focus ion beam (FIB Hitachi NX2000) at 30 kV for cutting and at 5 kV for final polish. A JEOL JEM

ARM 200F microscope equipped with a Cs-corrector at 200 kV was used to examine TEM/STEM specimens.

**Optical lithography for preparing Al Hall bar devices**

MBE-grown Al nanofilms were processed into Hall geometry by standard optical lithography. The width of the Hall bar and the center-to-center distance between the two voltage probes are 40 μm and 480 μm, respectively. Contact pads (200-nm-thick Al layer) made by standard optical lithography and metal evaporation were prepared for wire bonding. We used dilute AZ developer to avoid etching the Al nanofilm. The Al nanofilm devices were stable in air over a long period of time (over a year), possibly protected by the overlaying $AlO_x$ layer formed when the Al wafer was removed from the MBE chamber for *ex-situ* processing.

**Low-temperature electrical measurements**

Standard dc four-terminal *I-V* and sheet resistance measurements were performed on all Al Hall bar devices. A constant driving current was provided by a Keithley 2400 source meter and the voltage difference between voltage probes was measured using a Keithley 2000 multi-meter. Low-temperature experiments were performed in two cryo-free dilution refrigerators. Although the base temperature of our dilute refrigerators can be as low as 10 mK, we deliberately set the lowest temperature to 0.2 K (or 0.25 K) so that when dissipation occurs at a high critical current, Joule heating does not affect the set temperature (0.2 K or 0.25 K).

**SUPPLEMENTARY MATERIALS**

REFERENCE AND NOTES

1. D. B. Haviland, Y. Liu, A. M. Goldman, Onset of superconductivity in the two-dimensional limit. *Phys. Rev. Lett.* **62**, 2180-2183 (1989).
2. A. Yazdani, A. Kapitulnik, Superconducting-insulating transition in two-dimensional a-MoGe thin films. *Phys. Rev. Lett.* **74**, 3037-3040 (1994).
3. R. W. Cohen, B. Abeles, Superconductivity in granular aluminum films. *Phys. Rev.* **168**, 444-450 (1968).
4. J. J. Hauser, Enhancement of superconductivity in aluminum films. *Phys. Rev. B* **3**, 1611–1616 (1971).
5. R. B. Pettit, J. Silcox, Film structure and enhanced superconductivity in evaporated aluminum films. *Phys. Rev. B* **13**, 2865–2872 (1976).
6. V. N. Smolyaninova, K. Zander, T. Gresock, C. Jensen, J. C. Prestigiacomo, M. S. Osofsky, I. I. Smolyaninov, Using metamaterial nanoengineering to triple the superconducting critical temperature of bulk aluminum. *Sci. Rep.* **6**, 34140 (2016).
7. V. N. Smolyaninova, C. Jensen, W. Zimmerman, J. C. Prestigiacomo, M. S. Osofsky, H. Kim, N. Bassim, Z. Xing, M. M. Qazilbash, I. I. Smolyaninov, Enhanced superconductivity in aluminum-based hyperbolic metamaterials, *Sci. Rep.* **6**, 34140 (2016).
8. L. Grünhaupt, N. Maleeva, S. T. Skacel, M. Calvo, F. Levy-Bertrand, A. V. Ustinov, H. Rotzinger, A. Monfardini, G. Catelani, I. M. PopGrünhaupt, Loss mechanisms and quasiparticle dynamics in superconducting microwave resonators made of thin-film granular aluminum. *Phys. Rev. Lett.* **121**, 117001 (2018).
9. M. Strongin, R. S. Thompson, O. F. Kammerer, J. E. Crow, Destruction of superconductivity in disordered near-monolayer films. *Phys. Rev. B* **1**, 1078-1091 (1970).
10. D. Liu, W. Zhang, D. Mou, J. He, Y.-B. Ou, Q.-Y. Wang, Z. Li, L. Wang, L. Zhao, S. He, Y. Peng, X. Liu, C. Chen, L. Yu, G. Liu, X. Dong, J. Zhang, C. Chen, Z. Xu, J. Hu, X. Chen, X. Ma, Q. Xue, X. J. Zhou, Electronic origin of high-temperature superconductivity in single-layer FeSe superconductor. *Nat. Commun.* **3**, 931 (2012).
11. J. J. Lee, F. T. Schmitt, R. G. Moore, S. Johnston, Y.-T. Cui, W. Li, M. Yi, Z. K. Liu, M. Hashimoto, Y. Zhang, D. H. Lu, T. P. Devereaux, D.-H. Lee, Z.-X. Shen, Interfacial mode coupling as the origin of the enhancement of $T_c$ in FeSe films on $SrTiO_3$. *Nature* **515**, 245-248 (2014).
12. J.-F. Ge, Z.-L. Liu, C. Liu, C.-L. Gao, D. Qian, Q.-K. Xue, Y. Liu, J.-F. Jia, Superconductivity above 100 K in single-layer FeSe films on doped $SrTiO_3$. *Nat. Mater.* **14**, 285-289 (2015).
13. H. Ding, Y.-F. Lv, K. Zhao, W.-L. Wang, L. Wang, C.-L. Song, X. Chen, X.-C. Ma, Q.-K. Xue, High-temperature superconductivity in single-unit-cell FeSe films on Anatase $TiO_2$(001), *Phys. Rev. Lett.* **117**, 067001 (2016).
14. H.-M. Zhang, Y. Sun, W. Li, J.-P. Peng, C.-L. Song, Y. Xing, Q. Zhang, J. Guan, Z. Li, Y. Zhao, S. Ji, L. Wang, K. He, X. Chen, L. Gu, L. Ling, M. Tian, L. Li, X. C. Xie, J. Liu, H. Yang, Q.-K. Xue, J. Wang, Xucun. Ma, Detection of a superconducting phase in a two-atom layer of hexagonal

Ga film grown on semiconducting GaN(0001). *Phys. Rev. Lett.* **114**, 107003 (2015).
15. C. Zhang, F. Hao, G. Gao, X. Liu, C. Ma, Y. Lin, Y. Yin, X. Li, Enhanced superconductivity in TiO epitaxial thin films. *npj Quantum Materials* **2**, 2 (2017).
16. D. A. Rhodes, A. Jindal, N. F. Q. Yuan, Y. Jung, A. Antony, H. Wang, B. Kim, Y.-C. Chiu, T. Taniguchi, K. Watanabe, K. Barmak, L. Balicas, C. R. Dean, X. Qian, L. Fu, A. N. Pasupathy, J. Hone, Enhanced superconductivity in monolayer $T_d$-$MoTe_2$. *Nano Lett.* **21**, 2505–2511 (2021).
17. E. Khestanova, J. Birkbeck, M. Zhu, Y. Cao, G. L. Yu, D. Ghazaryan, J. Yin, H. Berger, L. Forró, T. Taniguchi, K. Watanabe, R. V. Gorbachev, A. Mishchenko, A. K. Geim, I. V. Grigorieva, Unusual suppression of the superconducting energy gap and critical temperature in atomically thin $NbSe_2$, *Nano Lett.* **18**, 2623-2629 (2018).
18. S. Mandal, S. Dutta, S. Basistha, I. Roy, J. Jesudasan, V. Bagwe, L. Benfatto, A. Thamizhavel, P. Raychaudhuri, Destruction of superconductivity through phase fluctuations in ultrathin $\alpha$-MoGe films. *Phys. Rev. B* **102**, 060501(R) (2020).
19. S. Giaremis, Ph. Komninou, Th. Karakostas, J. Kioseoglou, Ab Initio study of the electron–phonon coupling in ultrathin Al layers. *J. Low Temp. Phys.* **203**, 180–193 (2021).
20. W. van Weerdenburg, A. Kamlapure, E. HolmFyhn, X. Huang, N. P. E. van Mullekom, M. Steinbrecher, P. Krogstrup, J. Linder, A. A. Khajetoorians, Extreme enhancement of superconductivity in epitaxial aluminum near the monolayer limit. *Sci. Adv.* **9**, eadf5500 (2023).
21. J. M. Blatt, C. J. Thompson, Shape resonances in superconducting thin films. *Phys. Rev. Lett.* **10**, 332-334 (1963).
22. D. Valentinis, D. van der Marel, C. Berthod, Rise and fall of shape resonances in thin films of BCS superconductors, *Phys. Rev. B* **94**, 054516 (2016).
23. E. H. Hwang, S. Das Sarma, M. A. Stroscio, Role of confined phonons in thin-film superconductivity, *Phys. Rev. B* **61**, 8659-8662 (2000).
24. M. Shaukatullah, W. R. Storr, B. J. Hansen, Michael A. Gaynes, Design and optimization of pin fin heat sinks for low velocity applications, *IEEE Trans. Compon. Packaging. Manuf. Technol. Part A* **19**, 486-494 (1996).
25. H. Niwa, H. Yagi, H. Tsuchikawa, Stress distribution in an aluminum interconnect of very large scale integration. *J. Appl. Phys.* **68**, 328-333 (1990).
26. M. W. Knight, N. S. King, L. Liu, H. O. Everitt, P. Nordlander, N. J. Halas, Aluminum for plasmonics. *ACS Nano* **8**, 834–840 (2014).
27. P. K. Day, H. G. LeDuc, B. A. Mazin, A. Vayonakis, J. Zmuidzinas, A broadband superconducting detector suitable for use in large arrays. *Nature* **425**, 817-821 (2003).
28. G. De Simoni, F. Paolucci, P. Solinas, E. Strambini, F. Giazotto, Metallic supercurrent field-effect transistor. *Nat. Nanotechnol.* **13**, 802-805 (2018).
29. L. Grünhaupt, M. Spiecker, D. Gusenkova, N. Maleeva, S. T. Skacel, I. Takmakov, F. Valenti, P. Winkel, H. Rotzinger, W. Wernsdorfer, A. V. Ustinov, I. M. Pop, Granular aluminium as a superconducting material for high-impedance quantum circuits. *Nat. Mater.* **18**, 816–819 (2019).
30. S. Gustavsson, F. Yan, G. Catelani, J. Bylander, A. Kamal, J. Birenbaum, D. Hover, D. Rosenberg,

G. Samach, A. P. Sears, S. J. Weber, J. L. Yoder, J. Clarke, A. J. Kerman, F. Yoshihara, Y. Nakamura, T. P. Orlando, William D. Oliver, Suppressing relaxation in superconducting qubits by quasiparticle pumping. *Science* **354**, 1573-1577 (2016).

31. K. Epstein, A. M. Goldman, A. M. Kadin, Vortex-antivortex pair dissociation in two-dimensional superconductors. *Phys. Rev. Lett.* **47**, 534–537 (1981).
32. W. Zhao, Q. Wang, M. Liu, W. Zhang, Y. Wang, M. Chen, Y. Guo, K. He, X. Chen, Y. Wang, J. Wang, X. Xie, Q. Niu, L. Wang, X. Ma, J. K. Jain, M. H. W. Chan, Q.-K. Xue, Evidence for Berezinskii-Kosterlitz-Thouless transition in atomically flat two-dimensional Pb superconducting films. *Solid State Commun.* **165**, 59–63 (2013).
33. V. L. Berezinskii, Destruction of long range order in one-dimensional and two-dimensional systems having a continuous symmetry group. I. Classical systems. *Zh. Eksp. Teor. Fiz.* **59**, 907-920 (1970).
34. J. M. Kosterlitz, D. J. Thouless, Ordering, metastability and phase transitions in two-dimensional systems. *J. Phys. C* **6**, 1181-1203 (1973).
35. https://docs.quantumatk.com/tutorials/phonon_bs/phonon_bs.html
36. Z. Cheng, Y. R. Koh, H. Ahmad, R. Hu, J. Shi, M. E. Liao, Y. Wang, T. Bai, R. Li, E. Lee, E. A. Clinton, C. M. Matthews, Z. Engel, L. Yates, T. Luo, M. S. Goorsky, W. A. Doolittle, Z. Tian, P. E. Hopkins, S. Graham, Thermal conductance across harmonic-matched epitaxial Al-sapphire heterointerfaces. *Communications Physics* **3**, 115 (2020).
37. J. E. Herriman, B. Fultz, Phonon thermodynamics and elastic behavior of GaAs at high temperatures and pressures. *Phys. Rev. B* **101**, 214108 (2020).
38. Y. Liu, D. B. Haviland, B. Nease, A. M. Goldman, Insulator-to-superconductor transition in ultrathin films, *Phys. Rev. B* **47**, 5931 (1993).
39. Y. Ivry, C.-S. Kim, A. E. Dane, D. De Fazio, A. N. McCaughan, K. A. Sunter, Q. Zhao, K. K. Berggren, Universal scaling of the critical temperature for thin films near the superconducting-to-insulating transition, *Phys. Rev. B* **90**, 214515 (2014).
40. M. S. M. Minhaj, S. Meepagala, J. T. Chen, L. E. Wenger, Thickness dependence on the superconducting properties of thin Nb films, *Phys. Rev. B* **49**, 15235 (1994).
41. Y.-T. Fan, M.-C. Lo; C.-C. Wu, P.-Y. Chen, J.-S. Wu, C.-T. Liang, S.-D. Lin, Atomic-scale epitaxial aluminum film on GaAs substrate. *AIP Adv.* **7**, 075213 (2017).
42. Y. H. Tsai, Y.-H. Wu, Y.-Y. Ting, C.-C. Wu, J.-S. Wu, S.-D. Lin, Nano- to atomic-scale epitaxial aluminum films on Si substrate grown by molecular beam epitaxy. *AIP Adv.* **9**, 105001 (2019).

**Acknowledgments**

Funding: C. Panagopoulos acknowledges support from the National Research Foundation (NRF) Singapore Competitive Research Programme NRF-CRP21-2018-0001 and the Singapore Ministry of Education (MOE) Academic Research Fund Tier 3 Grant MOE2018-T3-1-002. C.-T.L., S.-D.L., H.L., and T.-R.C. acknowledge support from the National Science and Technology Council (NSTC), Taiwan.

# Supplementary Materials

**S1. Ratio of sheet resistance ($R$) to normal-state resistance ($R_N$) and sheet resistance of 3.5-nm-thick Al film on Si**

**S2. *I-V* curves of 3.5-nm-thick Al films on sapphire and on Si**

**S3. Critical temperature determined from $R = 0.9R_N$, $R = 0.5R_N$, $R = 0.1R_N$, and the BKT transitions**

**S4. Out-of-plane magneto-sheet-resistance and $H_{c2}(T)$ of 3.5 nm Al grown on GaAs, sapphire and Si**

**S5. *I-V* curves of Al nanofilms on GaAs and on sapphire**

**S6. AFM studies of 3.5 nm Al grown on GaAs, sapphire and Si**

**S7. In-plane radial scan of 3.5 nm Al grown on GaAs, sapphire and Si**

**S8. In-plane radial scan of Al films grown on GaAs of different thicknesses**

**S9. DFT calculations**

**S10. Residual-resistance ratio (*RRR*) of all Al films**

**S11. $\phi$ scans of thick Al films on GaAs, sapphire and Si**

**S12. Phonon DOS of bulk Al and GaAs, sapphire, and Si substrates**

**S13. Comparison with early work and absence of scaling**

**S14. In-plane and out-of-plane strain of 3.5 nm Al grown on GaAs, sapphire and Si**

## S1. Ratio of sheet resistance to normal-state resistance ($R_N$) and sheet resistance of 3.5-nm-thick Al film on Si

**A** **B** **C**

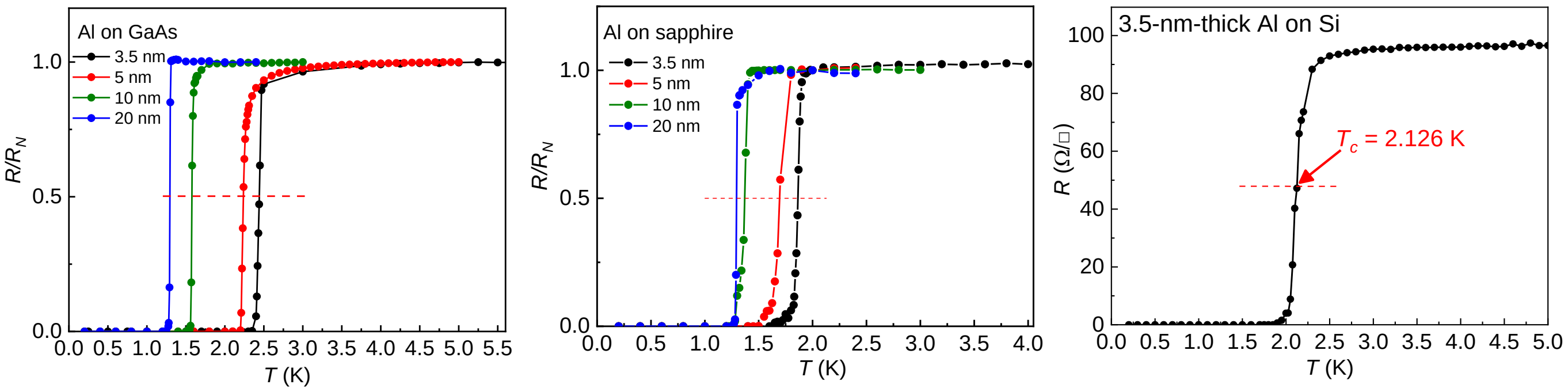

**Figure S1.** Ratio of sheet resistance to the normal-state sheet resistance as a function of temperature $R/R_N(T)$ for **(A)** Al nanofilms of different thicknesses $d$ grown on GaAs. **(B)** Al nanofilms of different thicknesses grown on sapphire. **(C)** $R(T)$ for 3.5-nm-thick Al grown on Si.

## S2. *I-V* curves of 3.5 nm Al films on sapphire and on Si

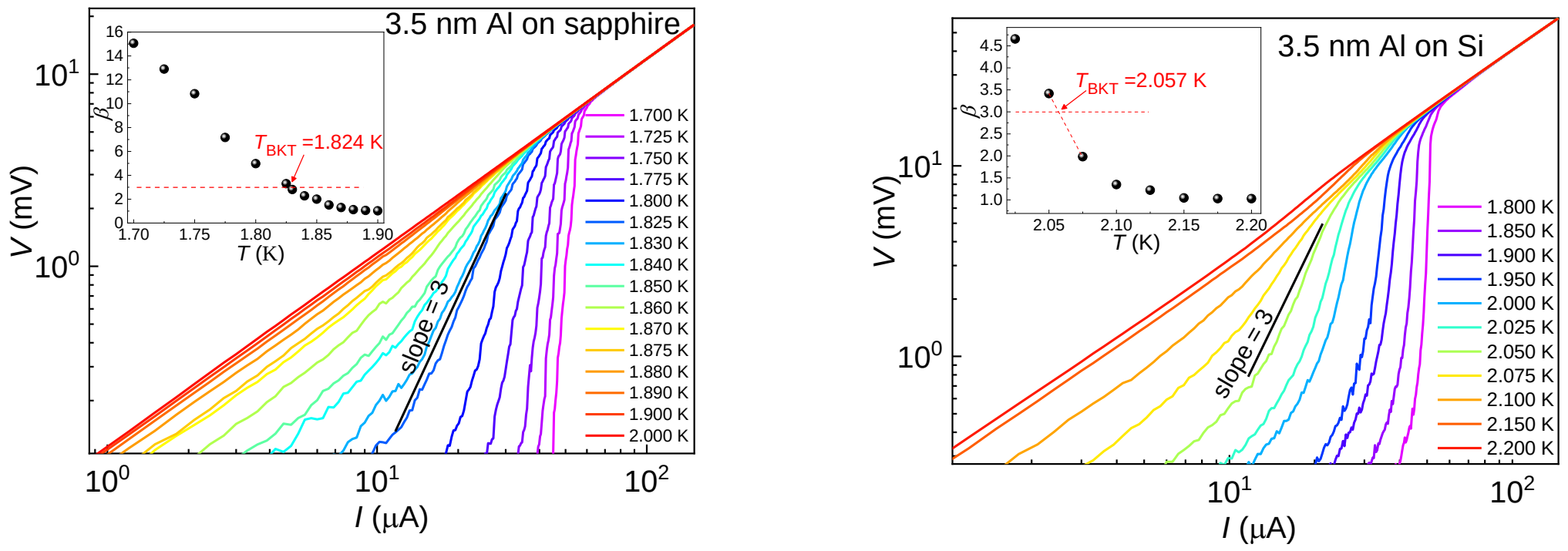

**Figure S2**. $I$-$V$ characteristics of the 3.5-nm-thick Al films grown on **(A)** sapphire and **(B)** on Si at various temperatures on a log-log scale. The insets show the exponent $\beta$ in the relation $V \sim I^{\beta}$ for the 3.5-nm-thick Al on **(A)** sapphire and **(B)** Si. The BKT transition temperatures $T_{BKT}$ are determined when $V \sim I^3$.

## S3. Critical temperature determined from $\boldsymbol{R = 0.9R_N}$, $\boldsymbol{R = 0.5R_N}$, $\boldsymbol{R = 0.1R_N}$, and the BKT transitions

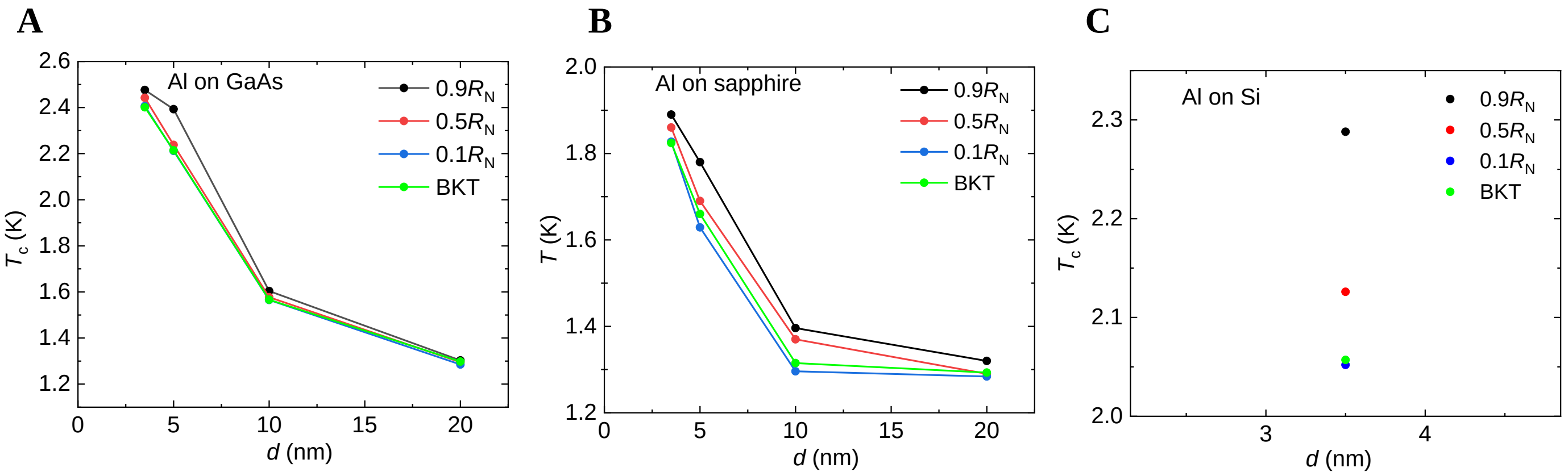

**Figure S3.** Critical temperature determined from $R = 0.9R_N$, $R = 0.5R_N$, $R = 0.1R_N$, and the BKT transitions for **(A)** Al on GaAs, **(B)** Al on sapphire at various thicknesses, and **c**, 3.5-nm-thick Al film on Si.

## S4. Out-of-plane magneto-sheet-resistance and $H_{c2}(T)$ of 3.5 nm Al grown on GaAs, sapphire and Si

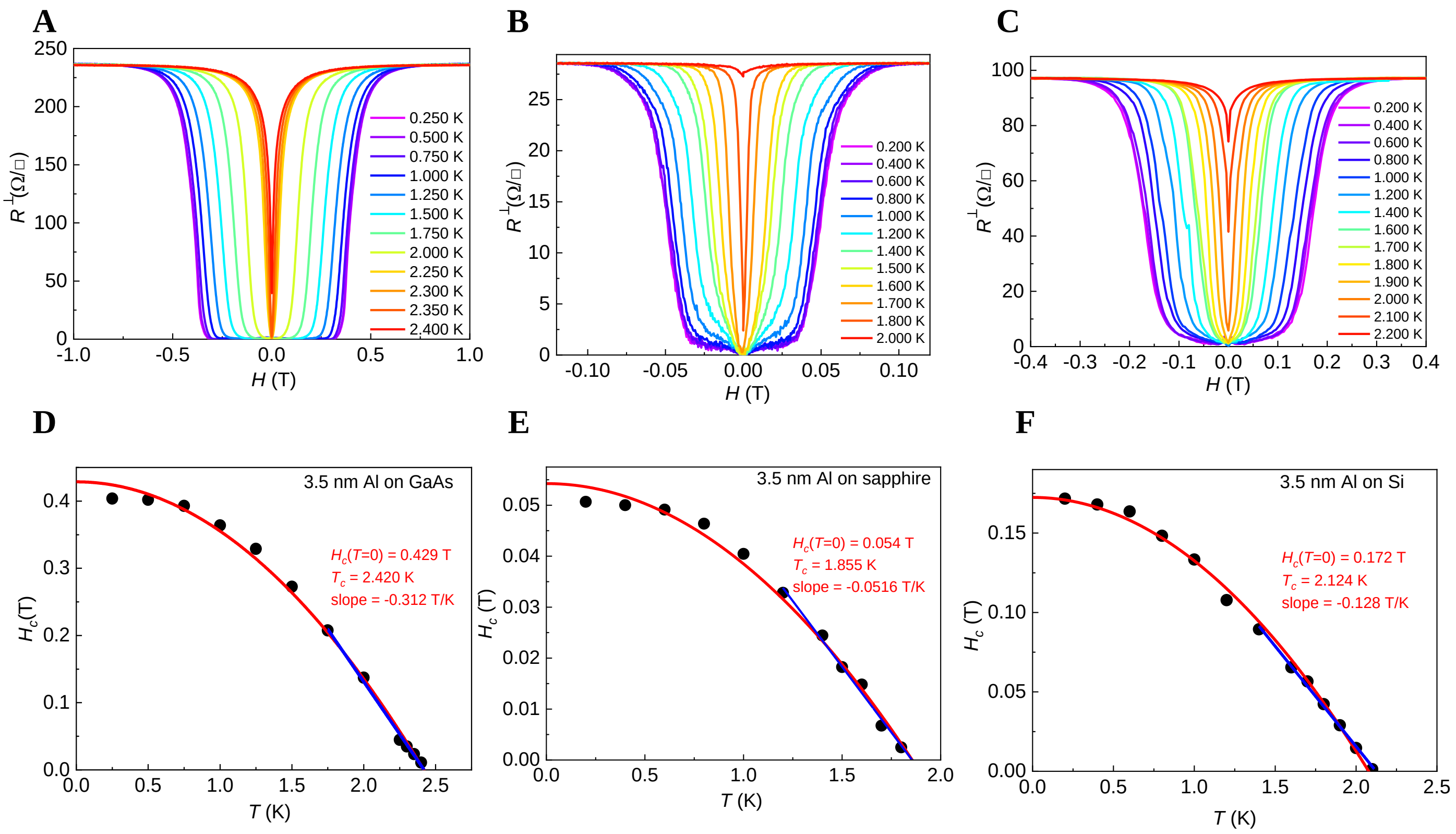

**Figure S4. Out-of-plane magneto-sheet-resistance $R^{\perp}(H)$ of the 3.5-nm-thick Al films at various temperatures.** (**A**) Al on GaAs, (**B**) Al on sapphire, and (**C**) Al on Si. (**D**) Critical magnetic field $H_c$ as a function of temperature for (**D**) Al on GaAs, (**E**) Al on sapphire, and (**F**) Al on Si. The fits (red curves) allow us to determine $H_c(T=0)$. The straight-line fits yield various slopes close to $T_c$.

## S5. *I-V* curves of Al nanofilms on GaAs and on sapphire

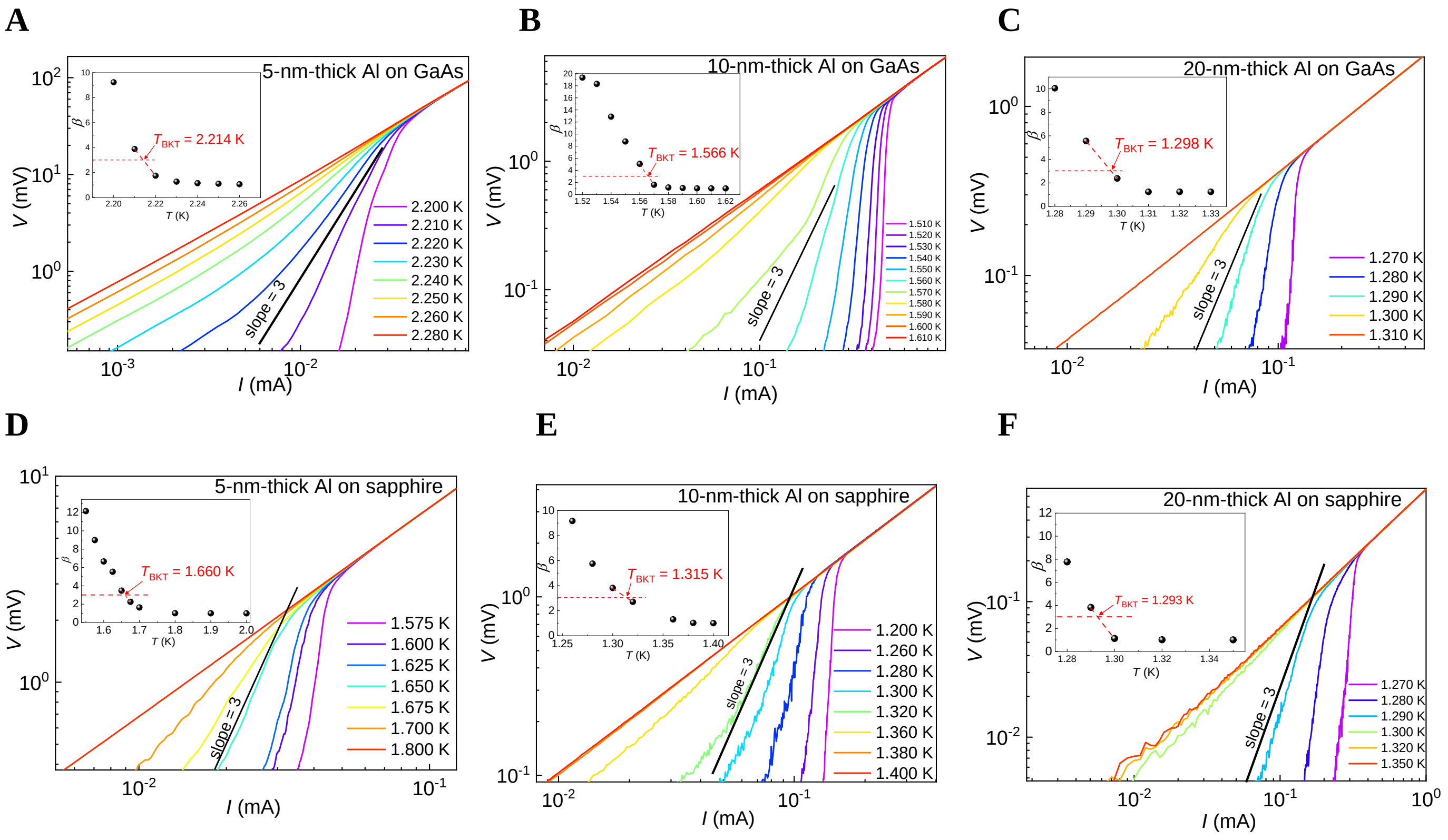

**Figure S5. (A)-(F) *I-V* characteristics of the Al films grown on GaAs and sapphire substrates at various temperatures on a log-log scale.** The insets show the exponent $\beta$ in the relation $V \sim I^{\beta}$. The BKT transition temperatures are determined when $V \sim I^3$ as shown in the insets.

### S6. AFM studies of 3.5 nm Al grown on GaAs, sapphire and Si

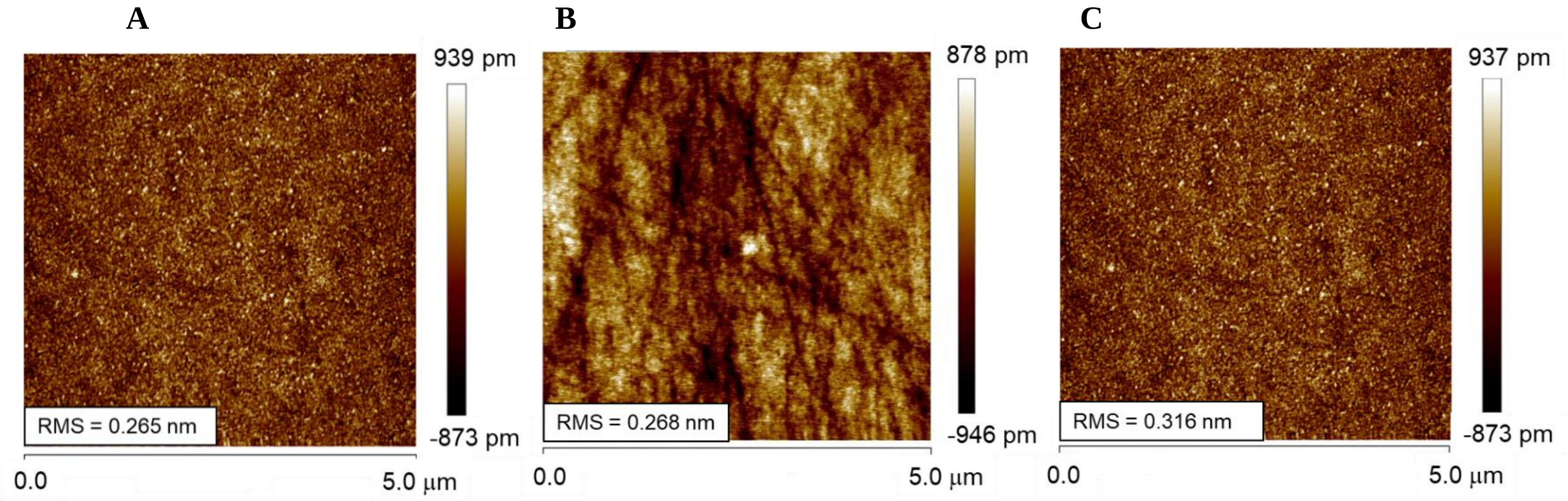

**Figure S6. AFM studies of the 3.5-nm-thick Al films** grown on (**A**) GaAs, (**B**) sapphire, and (**C**) Si, respectively.

## S7. In-plane radial scan of 3.5 nm Al grown on GaAs, sapphire and Si

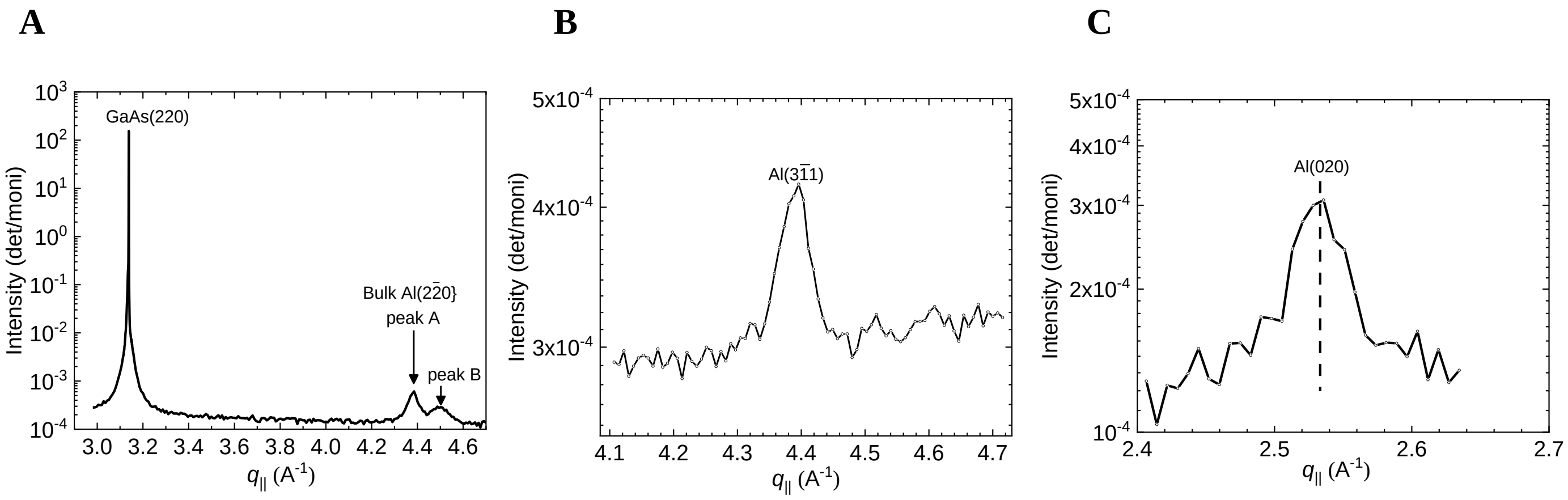

**Figure S7.** In-plane radial scan across **(A)** Al($2\bar{2}0$) reflection along GaAs[110] direction. **(B)** and **(C)** illustrate, respectively, the lateral scan across Al($3\bar{1}1$) reflection parallel to sapphire [$10\bar{1}0$] direction and across Al(020) peak parallel to the Si[$11\bar{2}$] direction.

## S8. In-plane radial scan of Al films grown on GaAs of different thicknesses

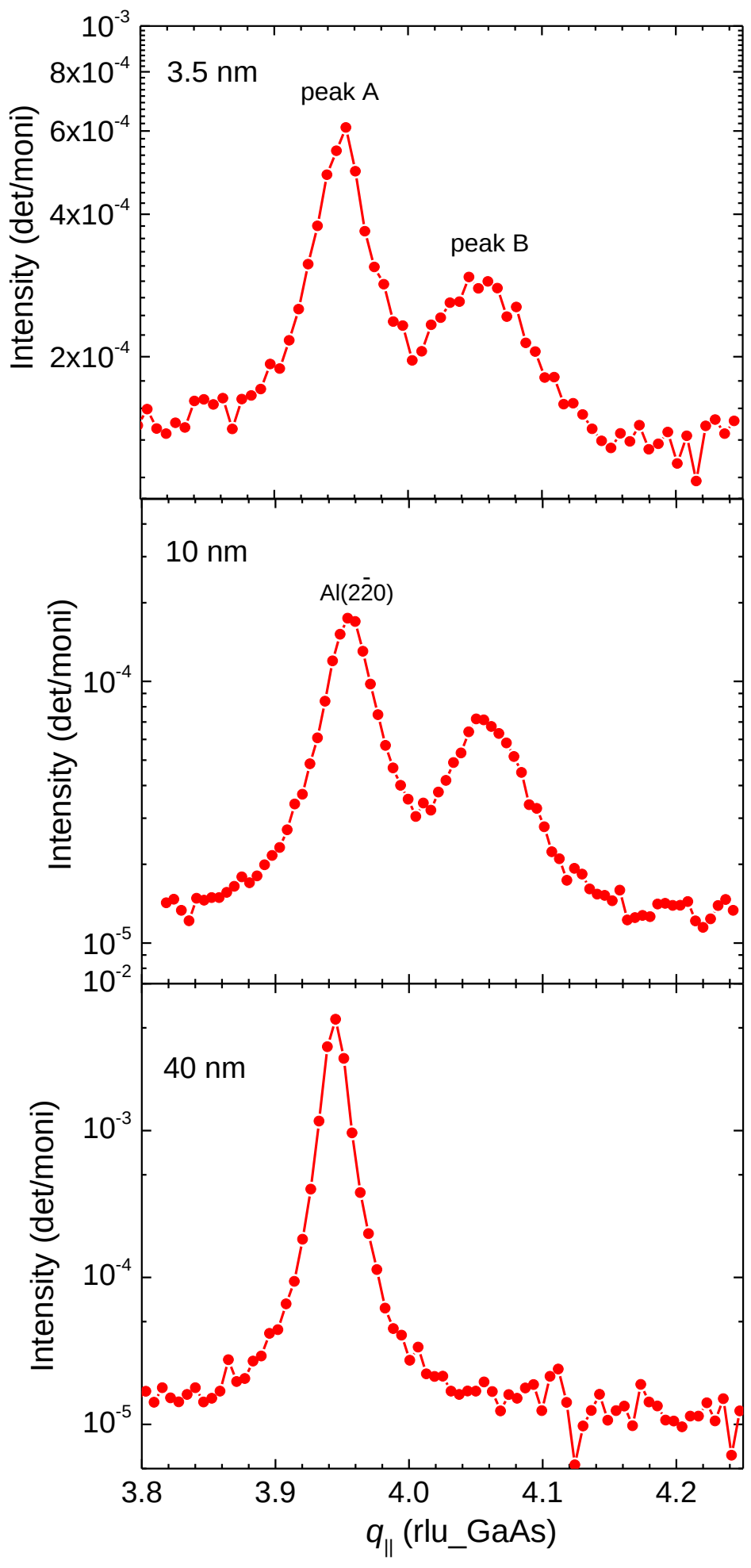

**Figure S8. In-plane radial scan across Al($2\bar{2}0$) reflection along GaAs[110] direction across peaks A and B for different Al film thicknesses.**

## S9. DFT calculations

### DFT calculations

DFT calculations were performed with the QUANTUM ESPRESSO package (*S1*, *S2*) using norm-conserving scalar-relativistic pseudopotentials (*S3*) within the GGA-PBE approximation (*S4*) and a plane waves cutoff of 80 Ry. The unshifted **k** mesh of $12 \times 12 \times 6$ and $12 \times 12 \times 1$ points for bulk and thin films, respectively. The structure was fully relaxed until the remanent forces were less than $0.136\ \mathrm{meV/Å}$. The electron-phonon calculations were computed within density functional perturbation theory (*S5*) on the $6 \times 6 \times 3$ and $6 \times 6 \times 1$ **q** mesh for bulk and thin films, respectively. The electron-phonon coupling strength $\lambda$ is given by

$$\lambda = \sum_{\mathbf{q}v} \lambda_{\mathbf{q}v} = \sum_{\mathbf{q}v} \frac{1}{\pi N(E_F)} \frac{\gamma_{\mathbf{q}v}}{\omega_{\mathbf{q}v}^2} = 2 \int \frac{\alpha^2 F(\omega)}{\omega} d\omega \tag{1}$$

where

$$\alpha^2 F(\omega) = \frac{1}{2\pi N(E_F)} \sum_{\mathbf{q}v} \frac{\gamma_{\mathbf{q}v}}{\omega_{\mathbf{q}v}} \delta\left(\omega - \omega_{\mathbf{q}v}\right) \tag{2}$$

is the isotropic Eliashberg spectral function, $v$ is the index of phonon modes, $N(E_F)$ is the electronic density of states (DOS) at the Fermi level, and $\gamma_{\mathbf{q}v}$ are phonon linewidths. We note that the phonon linewidths themselves are proportional to the electronic density of states; therefore (1) and (2), despite the appearances increase with $N(E_F)$. As one can see from Figure 9h, $N(E_F)$ increases with the decreasing thickness. This is indeed expected in the limit of strong confinement (*S6*). Thus, the theoretical enhancement of Tc can be attributed both to the phonon softening and increased phonon DOS as well as an increase in the electronic DOS. The critical temperature $T_c$ was estimated by the McMillan Allen-Dynes formula (*S7*, *S8*):

$$T_c = \frac{\omega_{\log}}{1.2} \exp\left[-\frac{1.04(1+\lambda)}{\lambda - \mu^*(1+0.62\lambda)}\right] \tag{3}$$

where

$$\omega_{\log} = \exp\left[\frac{2}{\lambda} \int \log(\omega) \frac{\alpha^2 F(\omega)}{\omega} d\omega\right] \tag{4}$$

is the logarithmic average frequency. The parameter $\mu^*$ is chosen to be 0.1.

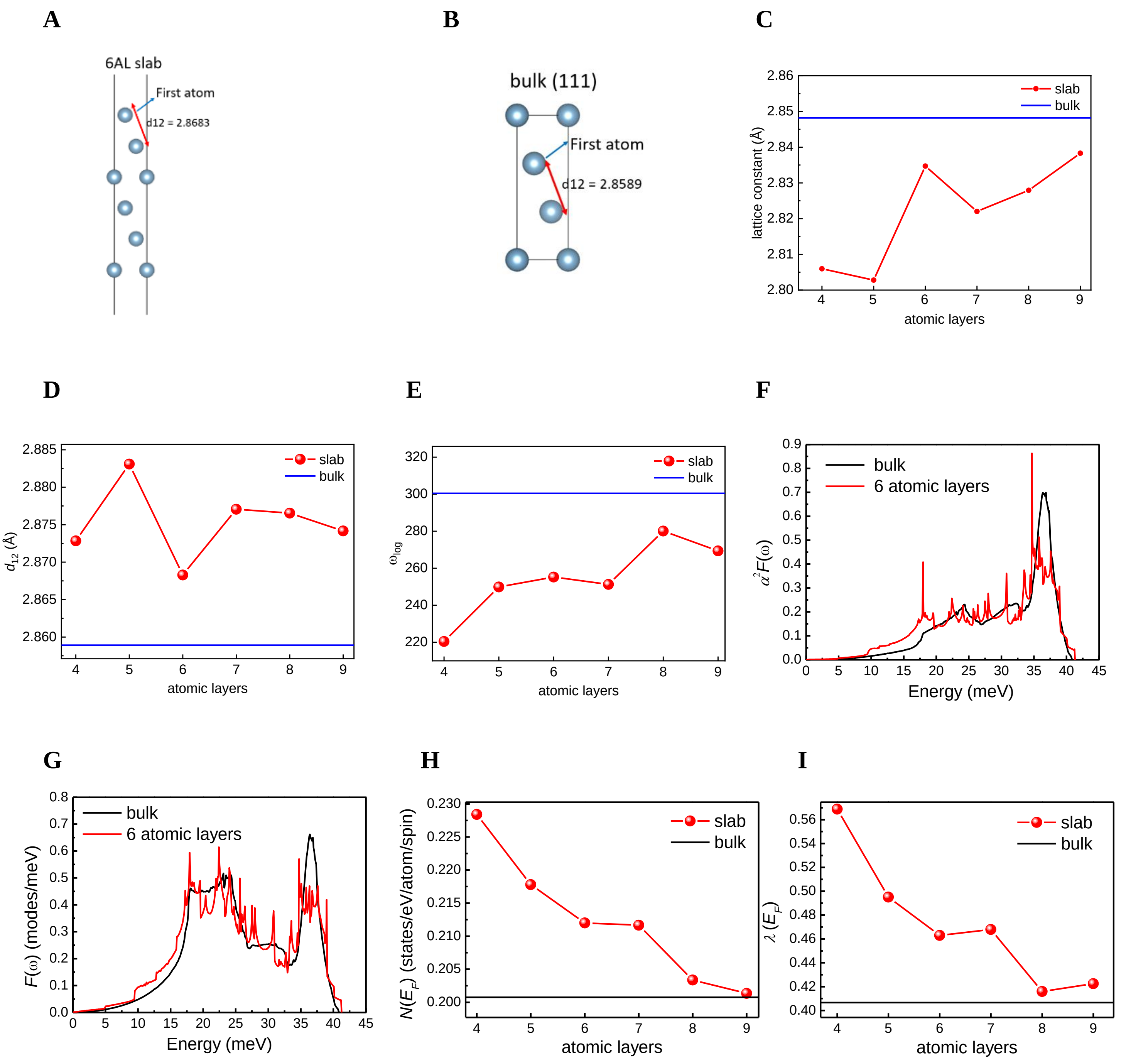

**Figure S9. (A) and (B) Models for performing substrate-free DFT calculations.** Thickness dependence of **(C)** the lattice constant, **(D)** the distance between the first and second atom from the top-most surface $d_{12}$ and **(E)** $\omega_{\log}$. We note that the $d_{12}$ profile is not monotonic because the lattice constant varies as well. Notably, $d_{12}$ for all thin films is larger than that in bulk Al. **(F)** Isotropic Eliashberg spectral function $\alpha^2F(\omega)$. **(G)** Phonon density of states $F(\omega)$. For comparison, the curves are normalized such that $\int F(\omega)d\omega = 9$. **(H)** thickness dependence of $N(E_F)$, **(I)** thickness dependence of $\lambda$.

## S10. Residual-resistance ratio (*RRR*) of all Al films

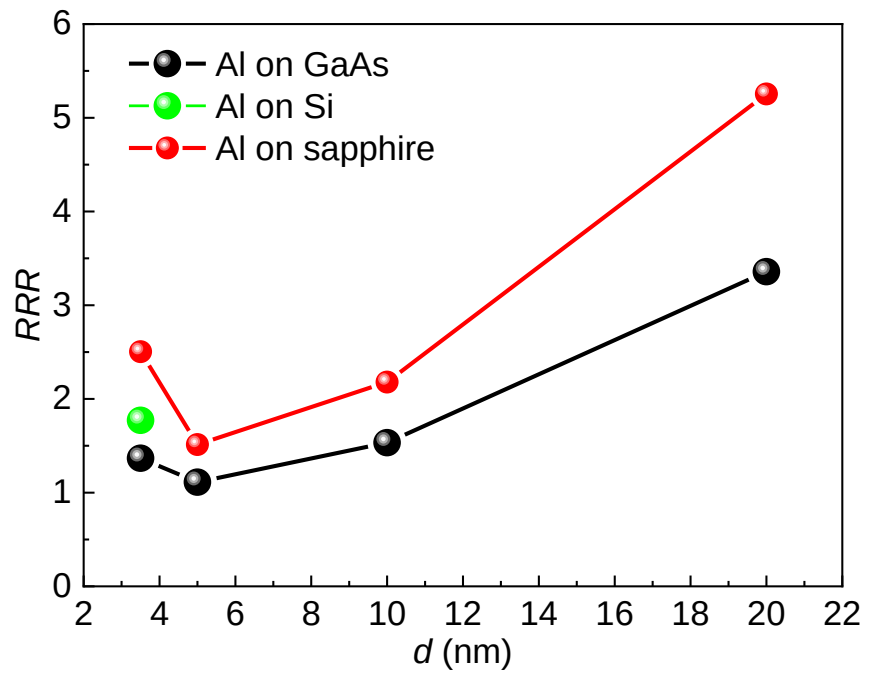

**Figure S10.** Residual-resistance ratio (*RRR*) $R_{300\ \mathrm{K}}/R_{\mathrm{N}}$ of the Al films on different substrates.

## S11. $\phi$ scans of thick Al films on GaAs, sapphire and Si

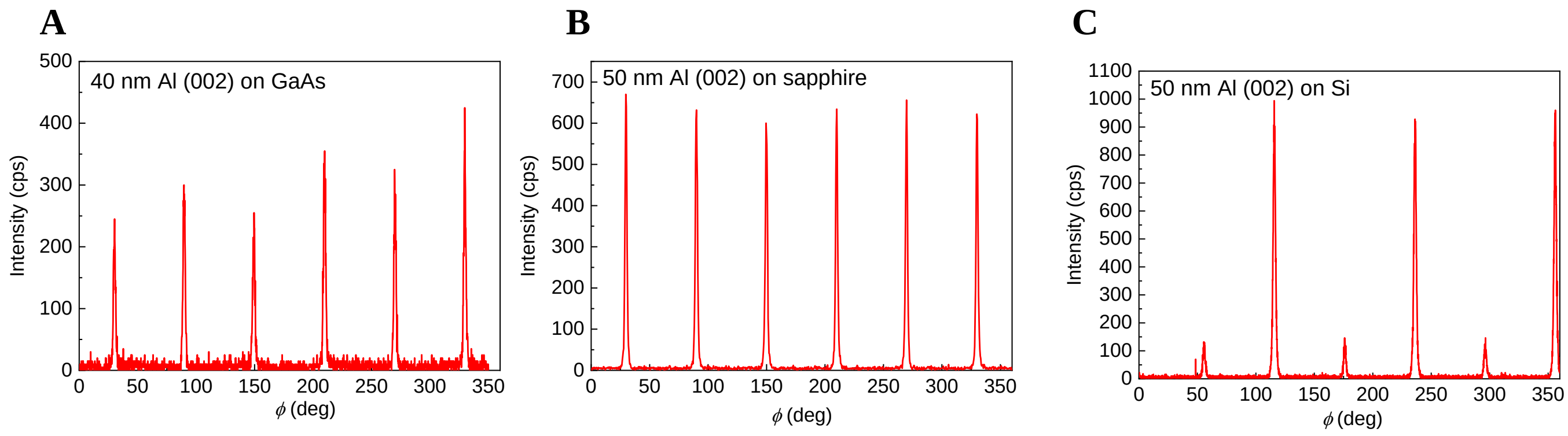

**Figure S11.** $\phi$ **scans of (A) 40-nm-thick Al on GaAs, (B) 50-nm-thick Al on sapphire and (C) 50-nm-thick Al on Si.**

## S12. Phonon DOS of bulk Al and GaAs, sapphire, and Si substrates

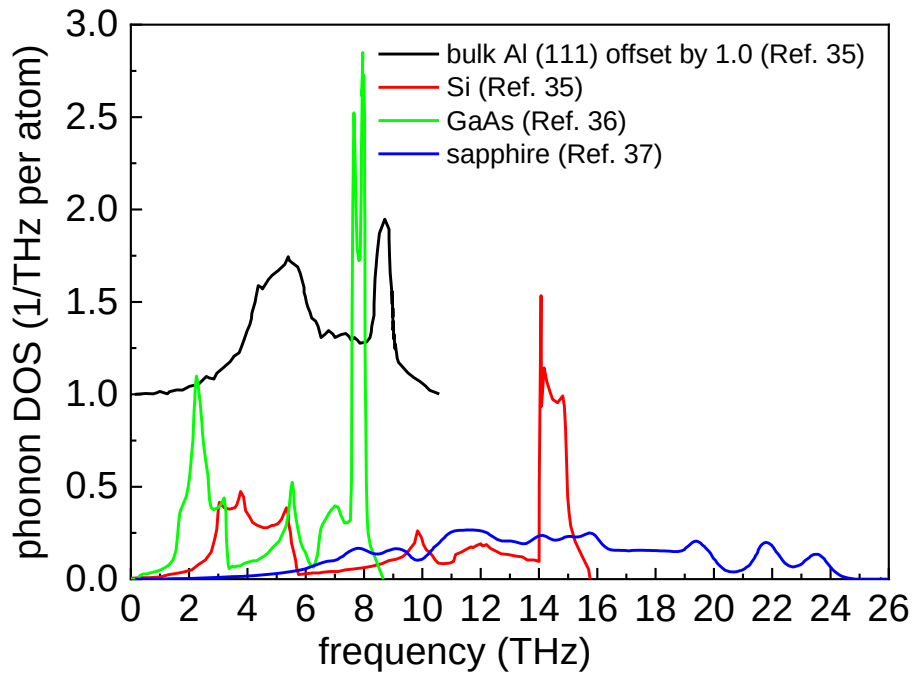

**Figure S12.** Phonon DOS (1/THz per atom) as a function of frequency (in THz) taken from Refs. 35-37. The curve corresponding to DOS of bulk Al (111) has been vertically offset by 1.0 for clarity.

## S13. Comparison with early work and absence of scaling

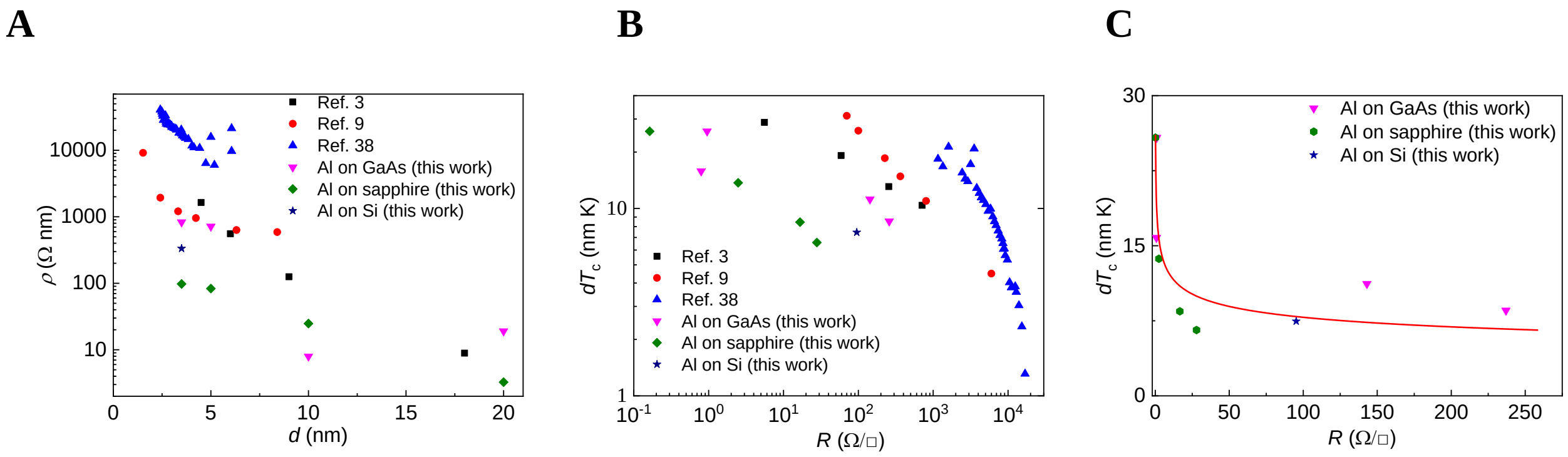

**Figure S13. (A)** Normal-state resistivity $\rho$ as a function of film thickness $d$. **(B)** product $dT_c$ as a function of normal-state sheet resistance $R$ taken on various samples. **(C)** Scaling fit $dT_c = AR^{-B}$ to our experimental data

**S14. In-plane and out-of-plane strain of 3.5 nm Al grown on GaAs, sapphire and Si**

| Standard bulk Al (Using 40 nm Al/GaAs value) | In-plane lattice constant | Out-of-plane lattice constant | | |
|---|---|---|---|---|
| | **2.864** | **7.017** | | |
| | | | **Lattice dilation calculation** | |
| | In-plane lattice constant | Out-of-plane lattice constant | **In-plane (%)** | **Out-of-plane (%)** |
| 3.5 nm Al on Si (111) | 2.858 | 7.11 | **-0.2** | **1.3** |
| | | | | |
| 3.5 nm Al on Sapphire (0001) | 2.864 | 7.011 | **0** | **-0.09** |
| | | | | |
| 3.5 nm Al on GaAs (001) (peak A) | 2.863 | within uncertainty | **-0.07** | **within uncertainty** |
| | | | | |

**Table S1**

**Supplementary References**

S1. P. Giannozzi, S. Baroni, N. Bonini, M. Calandra, R. Car, C. Cavazzoni, D. Ceresoli, G. L. Chiarotti, M. Cococcioni, I. Dabo, A. D. Corso, S. Fabris, G. Fratesi, S. de Gironcoli, R. Gebauer, U. Gerstmann, C. Gougoussis, A. Kokalj, M. Lazzeri, L. Martin-Samos, N. Marzari, F. Mauri, R. Mazzarello, S. Paolini, A. Pasquarello, L. Paulatto, C. Sbraccia, S. Scandolo, G. Sclauzero, A. P. Seitsonen, A. Smogunov, P. Umari, R. M. Wentzcovitch, QUANTUM ESPRESSO: a modular and open-source software project for quantum simulations of materials. *J. Phys. Condens. Matter* **21**, 395502 (2009).

S2. P. Giannozzi1, O. Andreussi, T. Brumme, O. Bunau, M. B. Nardelli, M. Calandra, R. Car, C. Cavazzoni, D. Ceresoli, M. Cococcioni, N. Colonna, I. Carnimeo, A. D. Corso, S. de Gironcoli, P. Delugas, R. A. DiStasio Jr, A. Ferretti, A. Floris, G. Fratesi, G. Fugallo, R. Gebauer, U. Gerstmann, F. Giustino, T. Gorni, J. Jia, M. Kawamura, H.-Y. Ko, A. Kokalj, E. Küçükbenli, M. Lazzeri, M. Marsili, N. Marzari, F. Mauri, N. L. Nguyen, H.-V. Nguyen, A. Otero-de-la-Roza, L. Paulatto, S. Poncé, D. Rocca, R. Sabatini, B. Santra, M. Schlipf, A. P. Seitsonen, A. Smogunov, I. Timrov, T. Thonhauser, P. Umari, N. Vast, X. Wu, S. Baroni, Advanced capabilities for materials modelling with Quantum ESPRESSO. *J. Phys. Condens. Matter* **29**,

465901 (2017).

S3. D. R. Hamann, Optimized norm-conserving Vanderbilt pseudopotentials. *Phys. Rev. B* **88**, 085117 (2013).

S4. J. P. Perdew, K. Burke, M. Ernzerhof, Generalized gradient approximation made simple. *Phys. Rev. Lett.* **77**, 3865–3868 (1996).

S5. S. Baroni, S. de Gironcoli, A. Dal Corso, P. Giannozzi, Phonons and related crystal properties from density-functional perturbation theory. *Rev. Mod. Phys.* **73**, 515–562 (2001).

S6. D. Valentinis, D. van der Marel, C. Berthod, Rise and fall of shape resonances in thin films of BCS superconductors, *Phys. Rev. B* **94**, 054516 (2016).

S7. W. L. McMillan, Transition temperature of strong-Coupled superconductors. *Phys. Rev.* **167**, 331–344 (1968).

S8. P. B. Allen, R. C. Dynes, Transition temperature of strong-coupled superconductors reanalyzed. *Phys. Rev. B* **12**, 905–922 (1975).